\documentclass{article}
\pdfoutput=1

\usepackage{amsmath, amssymb, amsthm, fullpage}
\usepackage{subfigure}

\date{}

\usepackage{graphicx}
\usepackage{subfigure}
\begin{document}

\newcommand{\BEAS}{\begin{eqnarray*}}
\newcommand{\EEAS}{\end{eqnarray*}}
\newcommand{\BEA}{\begin{eqnarray}}
\newcommand{\EEA}{\end{eqnarray}}
\newcommand{\BEQ}{\begin{equation}}
\newcommand{\EEQ}{\end{equation}}
\newcommand{\BIT}{\begin{itemize}}
\newcommand{\EIT}{\end{itemize}}
\newcommand{\R}{{\mathbf{R}}}
\newcommand{\x}{{\bf x}}
\newcommand{\ie}{{\it i.e.}}
\newcommand{\reals}{\mathbf{R}}
\newtheorem{lemma}{Lemma}
\newtheorem{theorem}{Theorem}
\newtheorem{proposition}{Proposition}
\newtheorem{corollary}{Corollary}


\title{Bidding for Representative Allocations for Display Advertising} 
\author{Arpita Ghosh\thanks{Yahoo! Research \texttt{\{arpita,mcafee,kpapi,sergei\}@yahoo-inc.com }} \and Preston McAfee\footnotemark[1] \and Kishore Papineni\footnotemark[1] \and Sergei Vassilvitskii\footnotemark[1]}

\maketitle

\begin{abstract}
Display advertising has traditionally been sold via guaranteed
contracts -- a guaranteed contract is a deal between a publisher and an
advertiser to allocate a certain number of impressions over a certain
period, for a pre-specified price per impression.  However, as spot
markets for display ads, such as the RightMedia Exchange, have grown
in prominence, the selection of advertisements to show on a given page
is increasingly being chosen based on price, using an auction. 
As the number of participants in the exchange grows, the price of an impressions becomes a signal of its value. 
This correlation between price and value means that a seller
implementing the contract through bidding should offer the contract
buyer a range of prices, and not just the cheapest impressions
necessary to fulfill  its demand. 

Implementing a contract using a range
of prices, is akin to creating a mutual fund of advertising impressions,
and requires {\em randomized bidding}.  We characterize what allocations
can be implemented with randomized bidding, namely those where the desired share obtained at each price is a non-increasing function of price. 
In addition, we provide a full
characterization of when a set of campaigns are compatible and how to
implement them with randomized bidding strategies.
\end{abstract}

\section{Introduction}

Display advertising --- showing graphical ads on regular web pages, as opposed to textual ads on search pages --- is approximately a \$24 billion business. There are two ways in which an advertiser looking to reach a specific audience (for example, 10 million males in California in July 2009) can buy such ad placements. One is the traditional method, where the advertiser enters into an agreement, called a guaranteed contract, directly with the publishers (owners of the webpages). Here, the publisher guarantees to deliver a prespecified number (10 million) of impressions matching the targeting requirements (male, from California) of the contract in the specified time frame (July 2009). 
The second is to participate in a spot market for display ads, such as the RightMedia Exchange, where advertisers can buy impressions one pageview at a time: every time a user loads a page with a spot for advertising, an auction is held where advertisers can bid for the opportunity to display a graphical ad to this user.  Both the guaranteed and spot markets for display advertising now thrive side-by-side. There is demand for guaranteed contracts from advertisers who want to hedge against future uncertainty of supply. For example, an advertiser who must reach a certain audience during a
critical period of time (e.g around a forthcoming product launch, such
as a movie release) may not want to risk the uncertainty of a spot
market; a guaranteed contract insures the publisher as well against fluctuations in
demand. At the same time, a spot market allows the advertisers to bid for specific opportunities, permitting very fine grained targeting based on user tracking. Currently, RightMedia runs over nine billion auctions for display ads everyday. 

How should a {\em publisher} decide which of her supply of impressions to allocate to her guaranteed contracts, and which to sell on the spot market?  One obvious solution is to fulfill the guaranteed demand first, and
then sell the remaining inventory on the spot market. However, spot
market prices are often quite different for two impressions that both
satisfy the targeting requirements of a guaranteed contract, since {\em different impressions have different value}. For example, the impressions from two users with identical demographics can have different value, based on different search behavior reflecting purchase intent
for one of the users, but not the other. Since advertisers on the spot market have access
to more tracking information about each user\footnote{For example, a
car dealership advertiser may observe that a particular user has been
to his webpage several times in the previous week, and may be willing
to bid more to show a car advertisement to induce a purchase. 
}, the resulting bids may be quite different for these two users.  Allocating impressions to guaranteed contracts first and selling the remainder on the spot market can therefore be highly suboptimal in terms of {\em revenue}, since two impressions that would fetch the same revenue from the guaranteed contract might fetch very different prices from the spot market\footnote {Consider the following toy example: suppose there are two opportunities,
the first of which would fetch 10 cents in the spot market, whereas
the second would fetch only $\epsilon$; both opportunities are equally
suitable for the guaranteed contract which wants just one
impression. Clearly, the first opportunity should be sold on the spot
market, and the second should be allocated to the guaranteed contract.}. 

On the other hand, simply buying the cheapest impressions on the spot market to satisfy guaranteed demand is not a good solution in terms of {\em fairness} to the guaranteed contracts, and leads to increasing short term revenue at the cost of long term satisfaction. 
As discussed above, impressions in online advertising have a {\em common value component}  
because advertisers generally have different information about a given user. This information (e.g. browsing history on an advertiser site) is typically relevant to {\em all} of the bidders, even though only {\em one} bidder may possess this information. 
In such settings, {\em price is a signal of value}---  in a model of valuations incorporating both common and private
values, the price converges to the true value of the item in the limit
as the number of bidders goes to infinity (\cite{Milgrom,Wilson}, see also \cite{Mcafee} for discussion).
On average, therefore, the price on the spot market is a good indicator of the value of the impression, and delivering cheapest impressions corresponds to delivering the lowest quality impressions to the guaranteed contract\footnote{While allocating the cheapest inventory to the guaranteed contracts is indeed revenue maximizing in the short term, in the long term the publisher runs the risk of losing the guaranteed advertisers by serving them the least valuable impressions.}. 

A publisher with access to both sources of demand thus faces a trade-off between revenue and fairness when deciding which impressions to allocate to the guaranteed contract; this trade-off is further compounded by the fact that the publisher 
typically does not have access to all the information that determines the value of a particular impression. Indeed, publishers are often the least well informed participants about the value of running an ad in front of a user. For example, when a user visits a politics site, Amazon (as an advertiser) can see that the user recently searched Amazon for an ipod, and Target (as an advertiser) can see they searched target.com for coffee mugs, but the publisher only knows the user visited the politics site. 
Furthermore, the exact nature of this trade-off is unknown to the publisher in advance, since it depends on the spot market bids which are revealed only {\em after} the advertising opportunity is placed on the spot market.

{\bf The publisher as a bidder.} To address the problem of unknown
spot market demand (\ie, the publisher would like to allocate the
opportunity to a bidder on the spot market {\em if} the bid is ``high
enough'', else to a guaranteed contract), the publisher acts, in effect,
as a {\em bidder} on behalf on the guaranteed contracts. That is, the
publisher now plays two roles: that of a seller, by placing his
opportunity on the spot market, and that of a bidding agent, bidding
on behalf of his guaranteed contracts. If the publisher's own bid
turns out to be highest among all bids, the opportunity is won and is
allocated to the guaranteed contract. Acting as a bidder allows
the publisher to probe the spot market and decide whether it is more
efficient to allocate the opportunity to an external bidder or to a
guaranteed contract. 

How should a publisher model the trade-off between fairness and revenue, and having decided on a trade-off, how should she place bids on the spot market?  An ideal solution is (a) easy to implement, (b) allows for a trade-off between the quality of
impressions delivered to the guaranteed contracts and short-term
revenues, 
and (c) is robust to the exact tradeoff chosen. In this work we show precisely when such an ideal solution exists and how it can be implemented. 

\subsection{Our Contributions}
In this paper, we provide an analytical framework to model the publisher's problem of how to fulfill guaranteed advance contracts
in a setting where there is an alternative spot market, and advertising opportunities have a common value component. 
We give a solution where the
publisher bids on behalf of its guaranteed contracts in the spot
market.  The solution consists of two components: an {\em allocation},
specifying the fraction of impressions at each price allocated to a
contract, and a {\em bidding strategy}, which specifies how to acquire
this allocation by bidding in an auction.  

The quality, or value, of an opportunity is measured by
its price \footnote{We emphasize that the assumption being made is {\em not} 
about price being a signal of value, but rather that impressions do have a
common value component -- given that impressions have a common value,
price reflecting value follows from the theorem of Milgrom~\cite{Milgrom}. This assumption is easily justifiable since it is commonly observed in practice.}.  A perfectly representative allocation is one which
consists of the same proportion of impressions at every price-- \ie,
a 
mix of high-quality and low quality impressions. The trade-off between revenue and fairness is modeled using a budget,
or {\em average target spend} constraint, for each advertiser's allocation: the publisher's choice of target spend reflects her trade-off between short-term revenue and quality of impressions for that advertiser (this must, of course, be large enough to ensure that the promised number of impressions satisfying the targeting constraints can be delivered.) Given a target spend \footnote{
We point out that we do not address the question of how to set target
spends, or the related problem of how to price guaranteed contracts to
begin with: {\em given} a target spend (presumably chosen based on
the price of guaranteed contracts and other considerations), we
propose a complete solution to the publisher' s problem.},
a maximally representative allocation is one which minimizes the
distance to the perfectly representative allocation, subject to the
budget constraint. We first show how to solve for a maximally
representative allocation, and then show how to implement such an
allocation by purchasing opportunities in an auction, using randomized
bidding strategies. 

{\bf Organization.} We start out with
the single contract case, where the publisher has just one existing
guaranteed contract, in Section \ref{s:single}; this case is enough to illustrate the idea of maximally representative allocations and
implementation via randomized bidding strategies. We move on to the more realistic case of multiple contracts
in Section \ref{s:mult}; we first prove a
result about which allocations can be implemented in an auction in a
decentralized fashion, and derive the corresponding decentralized
bidding strategies.  Next we solve for the optimal allocation when
there are multiple contracts. Finally, in Section \ref{sec:exp}, we
validate these strategies by simulating on data derived from real
world exchanges.

\subsubsection{Related Work}
The most relevant work is the literature on designing expressive auctions and clearing algorithms for online advertising  \cite{s1, s2, s4}. 
This literature does not address our problem for the following reason. While it is true that guaranteed contracts have coarse targeting relative to what is possible on the spot market, most advertisers with guaranteed contracts choose not to use all the expressiveness offered to them. 
Furthermore, the expressiveness offered does not include attributes like relevant browsing history on an advertiser site, which could increase the value of an impression to an advertiser, simply because the publisher {\em does not have} this information about the advertising opportunity. Even with extremely expressive auctions, 
 one might still want to adopt a mutual fund strategy to avoid the `insider trading' problem. That is, if some bidders possess good information about convertibility, others will still want to randomize their bidding strategy since bidding a constant price means always losing on some good impressions. Thus, our problem cannot be addressed by the use of more expressive auctions as in \cite{s4} --- the real problem is not lack of expressivity, but lack of information. 

Another area of research focuses on selecting the  optimal set of guaranteed contracts. In this line of work, Feige et al.~\cite{Feige} study the computational problem of choosing the set of guaranteed contracts to
maximize revenue. A similar problem is studied by in \cite{Muthu,
Bobby}. We do not address the problem of how to select the set of guaranteed contracts, but rather take them as given and address the problem of how to fulfill these contracts in the presence of competing demand from a spot market. 


\section{Single contract}
\label{s:single}
We first consider the simplest case: there is a single advertiser who
has a guaranteed contract with the publisher for delivering $d$
impressions. There are a total of $s \geq d$ advertising opportunities
which satisfy the targeting requirements of the contract. The
publisher can also sell these $s$ opportunities via auction in a spot
market to external bidders. The highest bid from the external bidders
comes from a {\em distribution} $F$, with density $f$, which we refer
to as the bid landscape. That is, for every unit of supply, the
highest bid from all external bidders,which we refer to as the price, is drawn i.i.d from the
distribution\footnote {Specifically, we do not consider adversarial
bid sequences; we also do not model the effect of the publisher's own
bids on others' bids.} $f$. (An example of such a density seen in a
real auction for advertising opportunities is shown in section
\ref{sec:exp}.) We assume that the supply $s$ and the bid landscape
$f$ are known to the publisher\footnote{Publishers typically have
access to data necessary to form estimates of these quantities; this is also discussed briefly in the conclusion}. Recall that the publisher
wants to decide how to allocate its inventory between the guaranteed
contract and the external bidders in the spot market. Due to penalties
as well as possible long term costs associated with underdelivering on
guaranteed contracts, we assume that the publisher wants to deliver
all $d$ impressions promised to the guaranteed contract.

An {\em allocation} $a(p)$ is defined as follows: $a(p)/s$ is the
proportion of opportunities at price $p$ purchased on behalf of the
guaranteed contract (the price is the highest (external)
bid for an opportunity.) That is, of the $sf(p)dp$ impressions available at price $p$, an allocation $a(p)$ buys a fraction $a(p)/s$ of these $sf(p)dp$ impressions, \ie, $a(p)f(p)dp$ impressions.  For example, a constant bid of $p^*$ means that for $p \leq p^*$, $a(p) = 1$ with the advertiser always winning the auction,  and for $p > p^*$, $a(p) = 0$ since the advertiser would never win.  

Generally, we will describe our solution in terms of
the allocation $a(p)/s$, which must integrate out to the total demand
$d$: a solution where $a(p)/s$ is larger for higher prices corresponds
to a solution where the guaranteed contract is allocated more
high-quality impressions. As another example, $a(p)/s = d/s$ is a perfectly representative allocation, integrating out to a total of $d$ impressions, and allocating the same fraction of impressions at every price point.  

Not every allocation can be purchased by bidding in an auction,
because of the inherent asymmetry in bidding-- a bid $b$ allows every
price below $b$ and rules out every price above; however, there is no
way to rule out prices {\em below} a certain value. That is, we can
choose to exclude high prices, but not low prices. Before
describing our solution, we state what kinds of allocations $a(p)/s$
can be purchased by bidding in an auction.

\begin{proposition}
\label{prop1}
A right-continuous allocation $a(p)/s$ can be implemented (in
expectation) by bidding in an auction if and only if $a(p_1) \geq
a(p_2)$ for $p_1 \leq p_2$.
\end{proposition}

\begin{proof}
  Given a right-continuous non-increasing allocation $\frac{a(p)}{s}$
  (that lies between 0 and 1), define $H(p) := 1 -\frac{a(p)}{s}.$ Let
  $p^* := \inf ~\{p: a(p) < s\}$. Then, $H$ is monotone non-decreasing
  and is right-continuous. Further, $H(p^*) = 0$ and $H(\infty) =
  1$. Thus, $H$ is a cumulative distribution function. We place bids
  drawn from $H$ (the probability of a strictly positive bid being
  $a(0)/s$).  Then the expected number of impressions won at price $p$
  is then exactly $a(p)/s$. Conversely, given that bids for the
  contract are drawn at random from a distribution $H$, the fraction
  of supply at price $p$ that is won by the contract is simply
  $1-H(p)$, the probability of its bid exceeding $p$.  Since $H$ is
  non-decreasing, the allocation (as a fraction of available supply at
  price $p$) must be non-increasing in $p$.
\end{proof}

Note that the distribution $H$ used to implement the allocation is a
different object from the bid landscape $f$ against which the
requisite allocation must be acquired-- in fact, it is completely
independent of $f$, and is specified only by the allocation
$a(p)/s$. That is, {\em given an allocation}, the bidding strategy
that implements the allocation in an auction is independent of the bid
landscape $f$ from which the competing bid is drawn.

\subsection{Maximally representative allocations} 
Ideally the advertiser with the guaranteed contract would like the
same proportion of impressions at every price $p$, \ie, $a(p)/s = d/s$
for all $p$.  (We ignore the possibility that the advertiser would
like a higher fraction of higher-priced impressions, since these
cannot be implemented according to Proposition \ref{prop1} above.)  However,
the publisher faces a trade-off between delivering high-quality
impressions to the guaranteed contract and allocating them to
bidders who value them highly on the spot market. We model this by introducing an average unit target spend $t$, which is the average
price of impressions allocated to the contract.  A smaller (bigger)
$t$ delivers more (less) cheap impressions. As we mentioned before, $t$ is part of the input problem, and may depend, for instance, on the price paid by the advertiser for the contract. 

Given a target spend, the {\em maximally representative} allocation is an
allocation $a(p)/s$ that is `closest' (according to some distance
measure) to the ideal allocation $d/s$, while respecting the target
spend constraint.  That is, it is the solution to the following
optimization problem:
\BEQ 
\begin{array}{cl} 
\label{o:maxrep}
  \inf_{a(\cdot)} & \int_p {\bf u}\left(\frac{a(p)}{s}, \frac{d}{s}\right)f(p)dp \\ [1ex]
  \mbox{s.t.} & \int_p a(p)f(p)dp = d \\ [1 ex]
  & \int_p pa(p)f(p)dp \leq td \\ [1 ex]
  & 0 \leq \frac{a(p)}{s} \leq 1. 
\end{array}  
\EEQ

The objective, ${\bf u}$, is a measure of the deviation of the proposed
fraction, $a(p)/s$, from the perfectly representative fraction,
$d/s$. In what follows, we will consider the $L_2$ measure
\[ 
{\bf u}\left(\frac{a(p)}{s}, \frac{d}{s}\right) = \frac{s}{2}\left(\frac{a(p)}{s} - \frac{d}{s}\right)^2 
\] 
as well as the Kullback-Leibler (KL) divergence
\[{\bf u}\left(\frac{a(p)}{s}, \frac{d}{s}\right) = \frac{a(p)}{s} \log
\frac{a(p)}{d}.\] 
Why the choice of KL and $L_2$ for ``closeness''?   Only Bregman divergences lead to a selection that is consistent, continuous, local, and transitive \cite{csiszar}. Further, in $R^n$ only least squares is scale- and translation- invariant, and for probability distributions only KL divergence is {\em statistical} \cite{csiszar}.  Indeed, KL is more appropriate in our setting. However, as least squares is more familar,  we discuss KL in Appendix \ref{sec:kl_appendix}.


The first constraint in (\ref{o:maxrep}) is simply that we must meet the target demand
$d$, buying $a(p)/s$ of the $sf(p)dp$ opportunities of price $p$.  The
second constraint is the {\em target spend} constraint: the total
spend (the spend on an impression of price $p$ is $p$) must not exceed
$td$, where $t$ is a target spend parameter (averaged per unit). As we will shortly see, the value of $t$ strongly affects the form of the solution.
Finally, the last constraint simply says that the proportion of
opportunities bought at price $p$, $a(p)/s$, must never go negative or exceed $1$.

{\bf Optimality conditions:}  Introduce Lagrange multipliers $\lambda_1$ and $\lambda_2$ for the
first and second constraints, and $\mu_1(p), \mu_2(p)$ for the two
inequalities in the last constraint. The Lagrangian is 
\BEAS
L &=& \int {\bf u}\left(\frac{a(p)}{s}, \frac{d}{s}\right) f(p)dp ~~+~~ \lambda_1\left(d - \int
a(p)f(p)dp\right) + \lambda_2\left(\int pa(p)f(p)dp - td\right) \\
&& ~+~ \int \mu_1(p)(-a(p)) f(p) dp + \int \mu_2(p)(a(p)-s)f(p)dp. 
\EEAS 
By the Euler-Lagrange conditions for optimality, the optimal solution must satisfy
\[ 
{\bf u}'\left(\frac{a(p)}{s}, \frac{d}{s}\right)= \lambda_1 - \lambda_2 p  +
\mu_1(p) - \mu_2(p),
\] 
where the multipliers $\mu$ satisfy $\mu_1(p), \mu_2(p) \geq 0$, and
each of these can be non-zero only if the corresponding constraint is
tight.

These optimality conditions, together with Proposition \ref{prop1}, give us the following:
\begin{proposition}
The maximally representative allocation for a single contract can be implemented by bidding in an auction for any {\em convex} distance measure $\bf{u}$. 
\end{proposition}
The proof follows from the fact that $\bf{u}'$ is increasing for convex $\bf{u}$. 

\subsubsection{$L_2$ utility}
\label{sec:l2_single}
In this subsection, we derive the optimal allocation when $\bf {u}$,
the distance measure, is the $L_2$ distance, and show how to
implement the optimal allocation using a randomized bidding
strategy. In this case the bidding strategy turns out to be very
simple: toss a coin to decide whether or not to bid, and, if bidding,
draw the bid value from a uniform distribution. The coin tossing
probability and the endpoints of the uniform distribution depend on
the demand and target spend values.

First we give the following result about the continuity of the optimal
allocation; this will be useful in deriving the values that
parameterize the optimal allocation. See Appendix \ref{sec:l2proof} for the proof. 

\begin{proposition} 
The optimal allocation $a(p)$ is continuous in $p$. 
\end{proposition}
Note that we do not assume \emph{a priori\/} that $a(\cdot)$ is
continuous; the \emph{optimal\/} allocation turns out to be
continuous. 

The optimality conditions, when $\bf{u}$ is the $L_2$ distance, are:
\[ 
\frac{a(p)}{s}- \frac{d}{s} = \lambda_1 - \lambda_2 p  +
\mu_1(p) - \mu_2(p),
\] 
where the nonnegative multipliers $\mu_1(p), \mu_2(p)$ can be non-zero
only if the corresponding constraints are tight.

The solution to the optimization problem (\ref{o:maxrep}) then takes
the following form: For $0 \leq p \leq p_{\min}$, $a(p)/s = 1$; for
$p_{\min} \leq p \leq p_{\max}$, $a(p)/s$ is proportional to $C - p$,
\ie, $a(p)/s = z(C-p)$; and for $p \geq p_{\max}$, $a(p)/s = 0$.  

To find the solution, we must find $p_{\min}, p_{\max}, z,$ and $C$.
Since $a(p)/s$ is continuous at $p_{\max}$, we must have $C =
p_{\max}$. By continuity at $p_{\min}$, if $p_{\min} > 0$ then
$z(C-p_{\min}) = 1$, so that $z = \frac{1}{p_{\max} - p_{\min}}$.
Thus, the optimal allocation $a(p)$ is always parametrized by two
quantities, and has one of the following two forms:
\begin{enumerate}
\item $a(p)/s = z(p_{max} - p)$ for $p \leq p_{\max}$ (and $0$
for $p \geq p_{\max}$). \\
\noindent
When the solution is parametrized by $z, p_{\max}$, these values must satisfy
\BEA
s\int_0^{p_{\max}}z(p_{\max}-p)f(p)dp &=& d  \label{e1}\\
s\int_0^{p_{\max}}zp(p_{\max}-p)f(p)dp &=& td \label{e2}
\EEA
Dividing (\ref{e1}) by (\ref{e2}) eliminates $z$ to give an equation
which is monotone in the variable $p_{\max}$, which can be solved, for instance, using binary search. 

\item $a(p)/s = 1$ for $p \leq p_{\min}$, and $a(p)/s = \frac{p_{\max} -
p}{p_{\max} - p_{\min}}$ for $p \leq p_{\max}$ (and $0$
thenceforth).\\
\noindent
When the solution is parametrized by $p_{\min}, p_{\max}$, these values must satisfy 
\begin{align}
\label{eqn:sc1} 
sF(p_{\min}) + \int_{p_{\min}}^{p_{\max}}s\frac{(p_{\max}-p)}{p_{\max} - p_{\min}}f(p)dp &= d\\ 
\label{eqn:sc2}
\int_0^{p_{\min}}spf(p)dp+ \int_{p_{\min}}^{p_{\max}}sp\frac{(p_{\max}-p)}{p_{\max} - p_{\min}}f(p)dp &= td.  
\end{align}
We show how to solve this system in Appendix \ref{sec:appendix_solving}.
\end{enumerate}
Note that the optimal allocation can be represented more compactly as 
\BEQ
\label{optl2}
\frac{a(p)}{s} = \min\{1, z(p_{max} - p)\}.
\EEQ

{\it Effect of varying target spend:} 
Varying the value of the target spend, $t$, while keeping the demand
$d$ fixed, leads to a tradeoff between representativeness and 
revenue from selling opportunities on the spot market, in the following way.
The minimum possible target spend, while meeting the target demand (in
expectation) is achieved by a solution where $p_{\min} = p_{\max}$ and
$a(p)/s = 1$ for $p$ less equal this value, and $0$ for greater. The
value of $p_{\min}$ is chosen so that
\[ 
\int_0^{p_{\min}} s f(p) dp = d \Rightarrow p_{\min} = F^{-1}(\frac{d}{s}).
\]  
This solution simply bids a flat value $p_{\min}$, and
corresponds to giving the cheapest possible inventory to the
advertiser, subject to meeting the demand constraint. This gives the
minimum possible total spend for this value of demand, of
\BEAS
\underbar{t} d &=& \int_0^{p_{\min}} s p f(p) dp = sF(p_{\min}) E[p| p \leq
p_{\min}] \\ ~ &=& d E[p| p \leq p_{\min}]
\EEAS
(Note that the maximum possible total spend that is maximally representative
while not overdelivering is
$R = \int pf(p)dp = dE[p] = d\bar{p}$.)

\begin{figure}[th]
\begin{center}
\includegraphics[bb = 0 0 2.5in 2.5in]{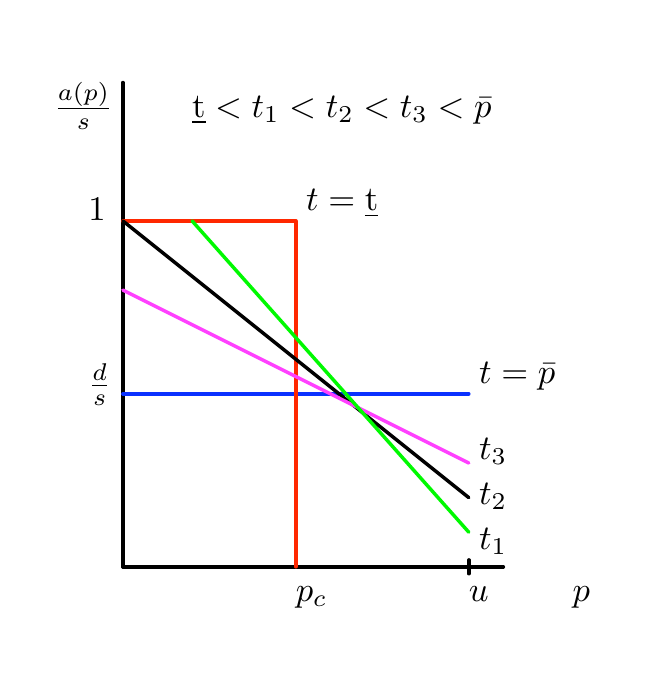}
\end{center}
\caption{Effect of target spend on $L_2$-optimal allocation \label{fig:l2alloc}}
\end{figure}

As the value of $t$ increases above $\underbar{t}$, $p_{\min}$
decreases and $p_{\max}$ increases, until we reach $p_{\min} = 0$, at
which point we move into the regime of the other optimal form, with $z
= 1$. As $t$ is increased further, $z$ decreases from $1$, and
$p_{\max}$ increases, until at the other extreme when the spend
constraint is essentially removed, the solution is $\frac{a(p)}{s} =
\frac{d}{s}$ for all $p$; \ie, a perfectly representative allocation
across price.  Thus the value of $t$ provides a dial by which to move
from the ``cheapest'' allocation to the perfectly representative
allocation. Figure \ref{fig:l2alloc} illustrates the effect of varying
target spend on the optimal allocation.

\subsection{Randomized bidding strategies}
The quantity $a(p)/s$ is an optimal {\em allocation}, \ie, a
recommendation to the publisher as to how much inventory to allocate
to a guaranteed contract at every price $p$.  However, recall that the
publisher needs to {\em acquire} this inventory on behalf of the
guaranteed contract by bidding in the spot market. The following
theorem shows how to do this when ${\bf u}$ is the $L_2$ distance.

\begin{theorem}
The optimal allocation for the $L_2$ distance measure can be
implemented (in expectation) in an auction by the following random
strategy: toss a coin to decide whether or not to bid, and if bidding,
draw the bid from a uniform distribution.
\end{theorem}
\begin{proof}
From (\ref{optl2}) that the optimal allocation can be represented as
\[ 
\frac{a(p)}{s} = \min\{1, z(p_{max} - p)\}.
\]
By Proposition \ref{prop1}, an allocation $\frac{a(p)}{s} = \min\{1, z(p_{max} - p)\}$ can be
implemented by bidding in an auction using the following randomized
bidding strategy: with probability $\min \{zp_{\max}, 1\}$, place a
bid drawn uniformly at random from the range $[\max\{p_{\max} -
\frac{1}{z},0\}, p_{\max}]$.
\end{proof}

The optimal allocation for KL-divergence decays exponentially with price, and the bidding strategy involves drawing bids from an exponential distribution; see Appendix \ref{sec:kl_appendix} for details.

\section{Multiple contracts} 
\label{s:mult}
We now study the more realistic case where the publisher needs to
fulfill multiple guaranteed contracts with different
advertisers. Specifically, suppose there are $m$ advertisers, with
demands $d_j$. As before, there are a total of $s \geq \sum d_j$ advertising
opportunities available to the publisher. \footnote {In general, not
all of these opportunities might be suitable for every contract; we do
not consider this here for clarity of presentation. However the same
ideas and methods can be applied in that most general case; the
results are also qualitatively similar.  } An allocation $a_j(p)/s$ is
the proportion of opportunities purchased on behalf of contract $j$ at
price $p$. Of course, the sum of these allocations cannot exceed 1 for
any $p$, which corresponds to acquiring all the supply at that price.

As in the single contract case, we are first interested in what
allocations $a_j(p)$ are implementable by bidding in an
auction. However, in addition to being implementable, we would like
allocations that satisfy an additional practical requirement,
explained below. Notice that the publisher, acting as a bidding agent, now
needs to acquire opportunities to implement the allocations for {\em
  each of the guaranteed contracts}. When an opportunity comes along,
therefore, the publisher needs to decide {\em which} of the contracts
(if any) will receive that opportunity.  There are two ways to do
this: the publisher submits {\em one} bid on behalf of all the
contracts; if this bid wins, the publisher then selects one amongst
the contracts to receive the opportunity.  Alternatively, the
publisher can submit one bid for {\em each} contract; the winning bid
then automatically decides which contract receives the opportunity.
We refer to the former as a {\bf centralized strategy} and the latter
as a {\bf decentralized strategy}.

There are situations where the publisher will need to choose the
winning advertiser {\em prior to} seeing the price, that is, the
highest bid from the spot market. For example, to reduce latency in
placing an advertisement, the auction mechanism may require that the
bids be accompanied by the advertisement (or its unique identifier).
A decentralized strategy automatically fulfills this requirement,
since there is one bid for each contract and the highest bid wins, so
that the choice of winning contract does not depend upon knowing the
price. In a centralized strategy, this requirement means that the
relative fractions won at price $p$, $a_i(p)/a_j(p)$, are {\em
independent of the price $p$}-- when this happens, the choice of advertiser can be made (by choosing at random with probability proportional to $a_j$) without knowing the price. 

 As before, we will be interested in implementing optimal (\ie,
maximally representative) allocations. For such an allocation, as we will
show in Section \ref{s:mult_optalloc}, if the relative fractions are
independent of the price, they can also be decentralized.  We will,
therefore, concentrate on characterizing allocations which can be
implemented via a decentralized strategy.

\subsection{Decentralization} 
In this section, we examine what allocations can be implemented via a
decentralized strategy. Note that it is not sufficient to simply use a
distribution $H_j = 1 - \frac{a_j(p)}{a_j(0)}$ as in Proposition
\ref{prop1}, since these contracts compete amongst each other as well.
Specifically, using the distribution $1 - \frac{a_j(p)}{a_j(0)}$ will
lead to too few opportunities being purchased for contract $j$, since
this distribution is designed to compete against $f$ alone,
rather than against $f$ {\em as well as} the other contracts.
We need to show how to choose distributions in such a way that lead to
a fraction $a_j(p)/s$ of opportunities being purchased for contract
$j$, for every $j = 1, \ldots, m$.

First, we argue that a decentralized strategy with given distributions
$H_j$ will lead to allocations that are non-increasing, as in the
single contract case. A decentralized implementation 
uses distributions $H_j$ to bid for impressions, \ie, it draws a
bid randomly from the distribution $H_j$ to place in the auction on
behalf of $j$-th contract.  Then, contract $j$ wins an impression at
price $p$ with probability
\[ 
a_j(p) = \int_p^\infty \left(\prod_{k \neq j} H_k(x)\right) h_j(x) dx, 
\] 
since to win, the bid for contract $j$ must be larger than $p$ {\em
  and} larger than the bids placed by each of the remaining $m-1$
contracts.  Since all the quantities in the integrand are nonnegative,
$a_j$ is non-increasing in $p$.

\medskip
\noindent
Now assume that $a_j$ are differentiable almost everywhere (a.e.) and non-increasing. Let
\[ 
\frac{A(p)}{s} := \sum_j  \frac{a_j(p)}{s}
\]
be the total fraction of opportunities at price $p$ that the publisher
needs to acquire.  Clearly, $a_j$ must be such that $A(p) \leq s,
~\forall p$.  Let $p* := \inf \{p: A(p) < s\}$. Now define
\BEQ 
\label{e:decentral}
H_j(p) := \left\{
\begin{array}{ll} e^{\int_p^\infty a'_j(x)/(s-A(x))dx} & ~~p > p^* \\ 
0 & ~~{\rm \mbox{else}}
\end{array}
\right.
\EEQ
Then, $H_j(p) \geq 0$ and is continuous. Since $a_j'(p)$ is non-increasing, $H_j(p)$ is monotone
non-decreasing. Further, $H(\infty) = 1$ and $H_j(p^*) = 0$. Thus, $H_j$
is a distribution function.  Now we verify that bidding according to
$H_j$ will result in the desired allocations: Note that
$$h_j(p) = \frac{d}{dp} H_j(p) = H_j(p) \frac{-a_j'(p)}{s-A(p)}$$
which implies
\[ 
\frac{A'(p)}{s-A(p)} = \sum_j \frac{a'_j(p)}{s-A(p)} = -\sum_j\frac{h_j(p)}{H_j(p)}, 
\] 
so that
\BEAS 
-\log(s-A(x))|_p^\infty &=& \int_p^\infty \frac{A'(x)}{s-A(x)}dx
= -\int_p^\infty \sum_j\frac{h_j(x)}{H_j(x)} dx 
= -\sum_j \log(H_j(x))|_p^\infty
\EEAS
or
\[ 
\log(s-A(p)) = \sum_j \log(H_j(p)) 
\] 
and hence $$\prod_k H_k(p) = s-A(p).$$ Then, the fraction of
impressions at $p$ that are won by contract $j$ is

\BEAS
\int_p^\infty \left(\prod_{k \neq j} H_k(x)\right) h_j(x) dx 
= \int_p^\infty \left(\prod_{k} H_k(x)\right) \frac{h_j(x)}{H_j(x)} dx 
= \int_p^\infty (s-A(x)) \frac{h_j(x)}{H_j(x)} dx 
= \int_p^\infty -a_j'(x) dx ~=~\frac{a_j(p)}{s} 
\EEAS 

\noindent
Thus, we constructed distribution functions $H_j(p)$ which implement
the given non-increasing (and a.e. differentiable) allocations
$a_j(p)$. If any $a_j$ is increasing at any point, the set of
campaigns cannot be decentralized. We summarize this in the following
theorem, whose special case for the single contract case is
Proposition \ref{prop1}:

\begin{theorem} 
A set of allocations $a_j(p)$ can be implemented in an auction via a
decentralized strategy if and only if each $a_j(p)$ is non-increasing
in $p$, and $\sum_j a_j(p)/s \leq 1$.
\end{theorem}

Having determined which allocations can be implemented by bidding in
an auction in a decentralized fashion, we turn to the question of
finding suitable allocations to implement. As in the single contract
case, we would like to implement allocations that are maximally
representative, given the spend constraints.

\subsection{Optimal allocation for multiple contracts}
\label{s:mult_optalloc}

As in the single contract case, every contract would ideally like an
equal proportion of opportunities at every price. However, every
contract has a per unit target spend which limits the fraction of opportunities
that can be purchased at higher prices. In addition to the target
spend, the allocation is also constrained by the fact that the total
fraction of opportunities bought at every price must not exceed
one. The maximally representative allocation is the allocation closest
to the ideal allocation that satisfies the target spend constraints, and
such that the collective allocation does not exceed the supply at any price.  That is, it is the
solution to the optimization problem below with the $L_2$ distance
measure in the objective.  We use $j$ to index the $m$ contracts.
\BEQ
\begin{array}{cl} 
\min & \frac{s}{2}\sum_{j=1}^m \int_p (\frac{a_j(p)}{s}- \frac{d_j}{s})^2 f(p)dp \\ [1 ex]
\mbox{s.t.} & \int_p a_j(p)f(p)dp = d_j \qquad \forall j \\ [1 ex]
& \int_p pa_j(p)f(p)dp \leq t_jd \qquad \forall j \\ [1 ex]
& a_j(p) \geq 0 \qquad \forall p,  j \\ [1 ex]
& \sum_{j=1}^m a_j(p)\leq s \qquad \forall p 
\end{array} 
\EEQ 
Observe that the allocations for individual contracts are coupled only
by the last constraint.

{\bf Optimality conditions:} Introduce Lagrange multipliers
$\lambda_1^j$ and $\lambda_2^j$ for the first and second constraints,
and $\mu_1^j(p), \mu_2(p)$ for the last two inequalities. The
optimality conditions are
\[ 
\frac{a_j(p)}{s}- \frac{d_j}{s} = \lambda^j_1 - \lambda^j_2 p +
\mu_1^j(p) - \mu_2(p),
\]
where $\lambda_2^j, \mu^j_1$ and $\mu_2$ must be nonnegative and can be non-zero only when
the corresponding constraint is tight.  Note that  $\mu_2$ is a contract-independent multiplier, corresponding to the coupling constraint. 

Suppose $a_j(p') \geq 0 ~\forall j$ and $\sum_j a_j(p') < s$ for some $p' >
0$. Then, $\mu_1^j(p') = \mu_2(p') = 0$. It follows that for $p > p'$,
$a_j(p) < a_j(p')$. Let $p^* = \inf ~\{p: \sum_j a_j(p) < s\}$. Then,
$\forall p \geq p^*$, each $a_j$ decays linearly with slope $\lambda_2^j$ until
it becomes 0. If $p^* = 0$, the solutions decouple, as $\mu_2(p)
\equiv 0$. In this case, we can solve for the $a_j$s independently of one
another. However, if $p^* > 0$, we have $\forall p < p^*$:
\[ 
\frac{a_j(p)}{s}- \frac{d_j}{s} = \lambda^j_1 - \lambda^j_2 p - \mu_2(p).
\]
Together with $\sum_j a_j(p) = s$, this implies
$$-\mu_2(p) = \frac{1}{m} - \frac{1}{ms} \sum_j d_j - \frac{1}{m} \sum_j \lambda_1^j + p \frac{1}{m} \sum_j \lambda_2^j.$$
Denoting $\bar{\lambda_2} := \frac{1}{m} \sum_j \lambda_2^j$, we see that 
$$\frac{a_j(p)}{s} = c_j - (\lambda_2^j - \bar{\lambda_2}) p, \qquad \forall p < p^*.$$ 

\begin{figure}[th]
\begin{center}
\includegraphics[bb=0 0 4.0in 2.0in, width=0.35\textwidth]{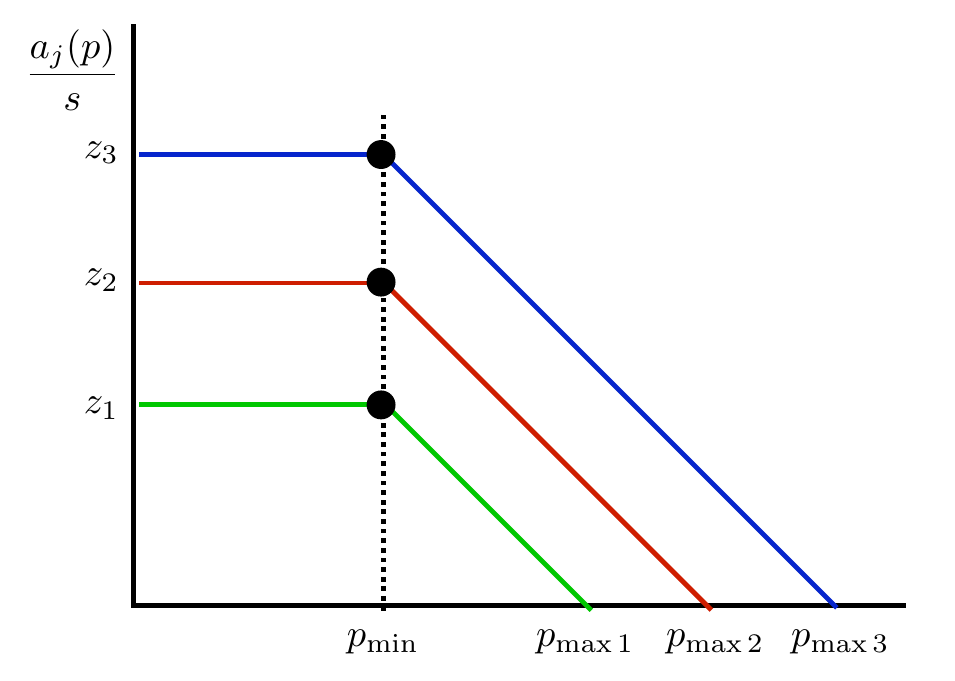}
\end{center}
\caption{Coupled decentralizable allocation.}
\label{fig:mult}
\end{figure}

Therefore, {\em at least one $a_j$ will have a positive slope} below
$p^*$ unless $\lambda_2^j = \lambda_2^k, ~\forall j, k$. That is, 
decentralization is not always guaranteed. In case the target spends
are such that $p^* > 0$ and $\lambda_2^j = \lambda_2^k, ~\forall j,
k$, the optimal allocations $a_j$ stay flat until $p^*$ and then decay
with identical slopes until each becomes 0, as shown in Figure \ref{fig:mult}.

Thus, the optimal allocation is decentralizable in two cases:
\begin{enumerate}\itemsep=-0.05in
\item $p^* = 0~$: The target spends are such that the solutions decouple. In this case the allocation for each contract is independent of the others; we solve for the parameters of each allocation as in Section \ref{sec:l2_single}.

\item $p^* > 0~$: The target spends are such that, for all $j, k, ~\frac{a_j(p)}{a_k(p)}$ is independent of $p$. In this case we need to solve for the common slope and $p_{\min}$, and the contract specific values $p^j_{\max}$, which together determine the allocation. This can be done using, for instance, Newton's method. 
\end{enumerate}

When the target spends are such that the allocation is not
decentralizable, the vector of target spends can be increased to reach
a decentralizable allocation\footnote{ We do not investigate the
approach of finding the best suboptimal allocation that can be
decentralized, \ie, an approximately optimal decentralizable
allocation, in this paper.}. One way is to scale up the target spends
uniformly until they are large enough to admit a separable solution;
this has the advantage of preserving the relative ratios of target
spends. The minimum multiplier which renders the allocation
decentralizable can be found numerically, using for instance binary
search.

\begin{figure}[t]
\begin{center}
\includegraphics[bb=0 0 8.0in 5.5in, width=0.35\textwidth]{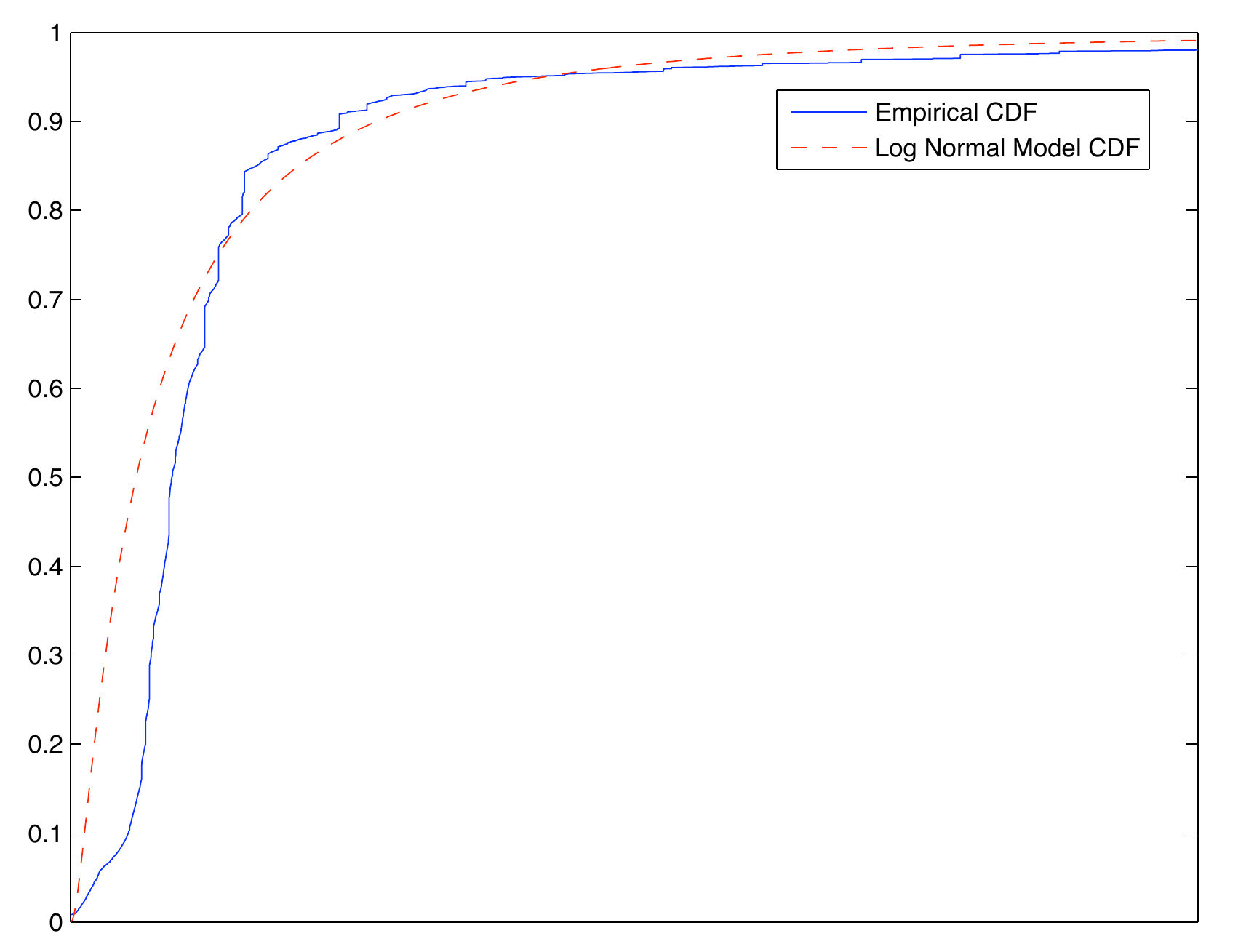}
\end{center}
\caption{Empirical cumulative density function of the bids on the exchange and a log-normal fit to the distribution.}
\label{fig:cdf}
\end{figure}
\vspace{-0.10in}
\section{Experimental Validation}
\label{sec:exp}
Our algorithms for obtaining representative allocations are
randomized, and all of the results are derived in expectation. In this
section we simulate the performance of the algorithms, and verify that
the randomization does not lead to under-delivery for a realistic
choice of bid landscape $f$.

To simulate the bid landscape, data was collected from live
auctions conducted by the RightMedia exchange. RightMedia runs the
largest spot market for display advertising, with billions of auctions
daily. Winning bids were collected from approximately 400,000
auctions over the course of a day for a specific publisher. The cdf of
the empirical bid distribution is plotted in Figure
\ref{fig:cdf}. (The scale on the x-axis is omitted for privacy
concerns.)

The empirical distribution is well approximated by a log-normal
distribution, as seen in Figure \ref{fig:cdf}. For the experimental
evaluation, we, therefore, draw bids from a log-normal
distribution. The mean of the distribution is set to $0$, and the
variance parameter is changed to investigate the sensitivity of our
algorithms to the variance of the bid landscape.

\begin{figure}[t]
\centering
\subfigure[Allocation achieved for different exchange distributions with ideal allocation of $0.25$, $0.5$ and $0.75$ fraction of the total.]{
\includegraphics[bb=0 0 8.0in 5.5in, width=0.45\textwidth]{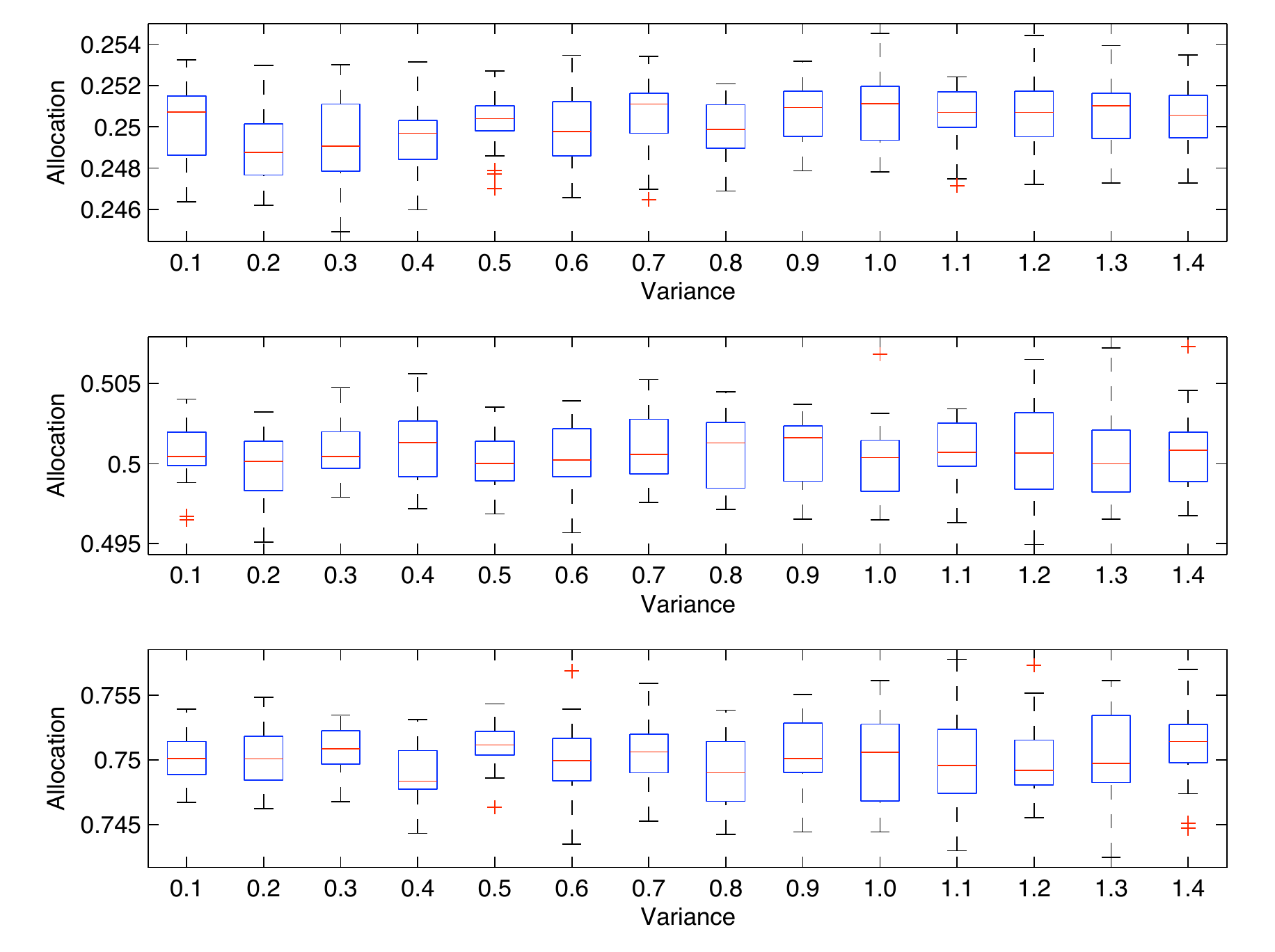}
\label{fig:allocation}
}
\subfigure[Target spend achieved for different exchange distributions with ideal target $0.25$, $0.5$ and $0.75$ of the maximum feasible spend.]{
\includegraphics[bb=0 0 8.0in 5.5in, width=0.45\textwidth]{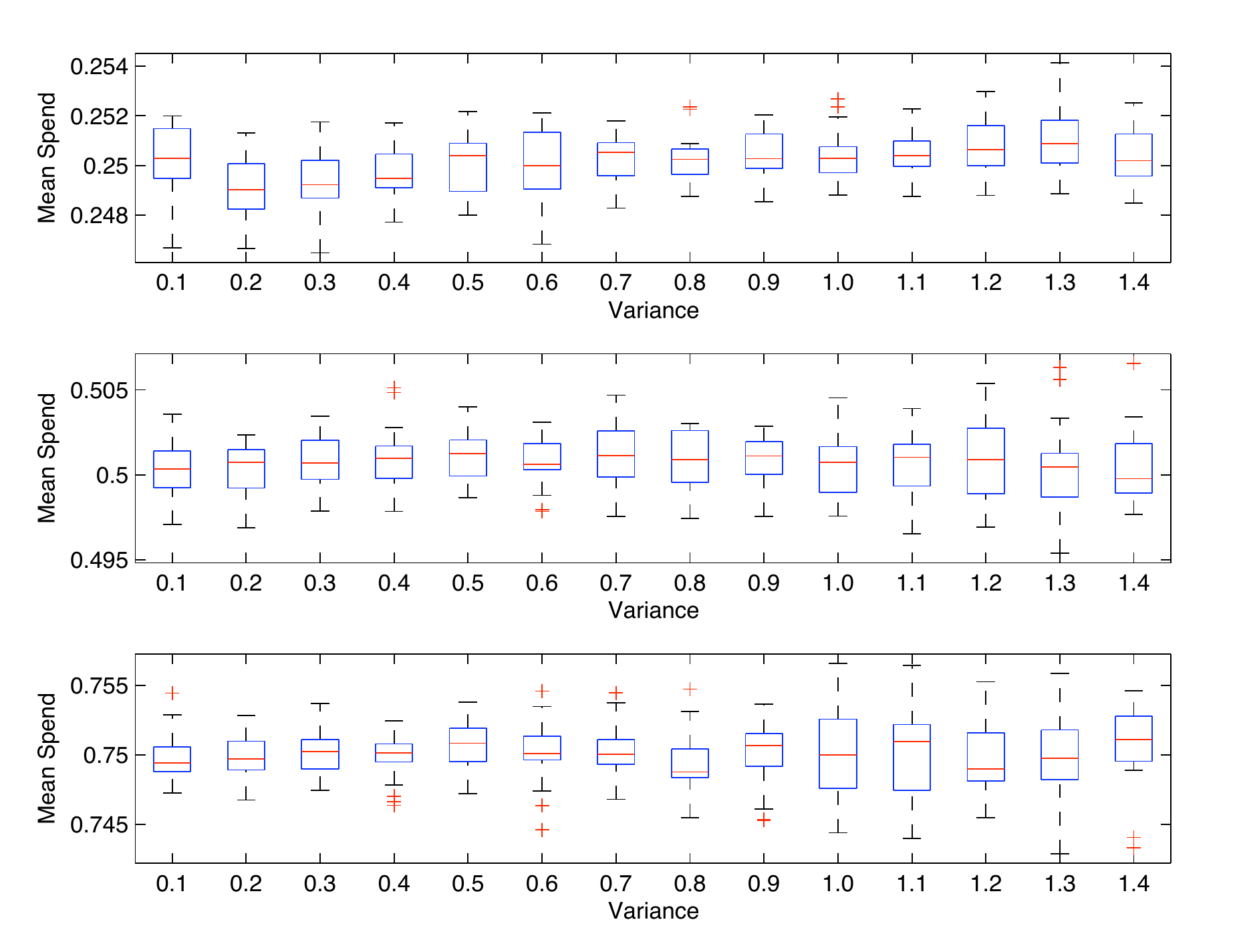}
\label{fig:targetSpend}
}
\caption{ Experiment evaluation}
\end{figure}

To study the effectiveness of the algorithm in winning the right
number of impressions on the exchange, we fix the target fraction,
$\frac{d}{s}$ at $0.25, 0.5$ and $0.75$, and compute the $p_{\max}$
necessary to achieve the allocation, yet minimize the total spend. For
each setting of the variance of the exchange distribution, we run 15
trials, each with 10,000 auctions total. The results are plotted in
Figure \ref{fig:allocation}.

We perform a similar experiment to investigate the dependency of the
target spend on the variance of the bid distribution. In this case, we
fix the allocation fraction to $0.8$ and target spend to $0.25\mu$,
$0.5\mu$ and $0.75\mu$, where $\mu = \int_p p f(p) dp$ is the maximum
achievable target spend. For each setting of the variance parameter we
run 15 trials each with 10,000 auctions. The results are plotted in
Figure \ref{fig:targetSpend}.

%

In both simulations, the specific $p_{\min}$ and $p_{\max}$ for each
variance setting vary greatly to achieve the desired allocation and
target spend. However, the changes in the resulting $\frac{a}{s}$ and
$\frac{d}{s}$ themselves are minimal -- the algorithm rarely
underdelivers or under/overspends by more than 1\%; it is robust to
variations in the variance of the underlying bid distributions.

\vspace{-0.05in}
\section{Conclusion}


Moving guaranteed contracts into an exchange environment presents a variety of challenges for a publisher.  Randomized bidding is a useful compromise between minimizing the cost and maximizing the quality of guaranteed contracts.  It is akin to the mutual fund strategy common in the capital asset pricing model.  We provide a readily computable solution for synchronizing an arbitrary number of guaranteed campaigns in an exchange environment.  Moreover, the solution we detail appears stable with real data.

There are many interesting directions for further research. We assumed
throughout that the supply is known to the publisher. A more realistic
model assumes either an unknown or a stochastic supply (a strawman
solution is to use algorithms in this paper using a lower bound on the
supply in place of $s$). Another interesting avenue is analyzing the
strategic behavior by other bidders on the spot market in response to
such randomized bidding strategies.

\bibliographystyle{plain}
\bibliography{randbids}

\section*{Appendix}
\appendix
\section{Solving for $p_{\min}$ and $p_{\max}$}
\label{sec:appendix_solving}
Calling $p_{\min}$ and $p_{\max}$ $x$ and $y$ respectively, we want to solve the system of equations:
$$f(y,x) = \frac{d}{s}\left[ \begin{array}{c} 1 \\ t \end{array} \right],$$
where 
\begin{align*}
f(y,x) = \left[\begin{array}{c}
 F(x) + \int_{x}^y \frac{y-p}{p-x}f(p) dp \\ [1ex]
 \int_{0}^x p f(p) dp + \int_{x}^y p \frac{y-p}{y-x} f(p) dp
\end{array}
  \right].
\end{align*}
We will show that the derivative matrix is invertible so we can use Newton's method to converge to the solution. 

The derivative of f is:
\begin{align*}
f' &= \left[
\begin{array}{cc}
\int_x^y \frac{p-x}{(y-x)^2}f(p)dp  & \int_x^y \frac{y-p}{(y-x)^2} f(p) dp
\\ [1 ex]
\int_x^y p \frac{p-x}{(y-x)^2} f(p) dp
&
\int_x^y p \frac{y-p}{(y-x)^2}f(p) dp 
\end{array}
\right]
\\[1 ex]
& =\frac{1}{(y-x)^2} \left[
\begin{array}{cc}
\int_x^y (p-x) f(p) dp 
&
\int_x^y (y-p) f(p) dp 
\\ [1ex]
\int_x^yp(p-x) f(p) dp 
&
\int_x^y p (y-p) f(p) dp 
\end{array}
\right]
\\[1 ex]
&= \frac{F(y) - F(x)}{(y-x)^2} \left[
\begin{array}{cc}
Ep-x
&
y-Ep
\\[1ex]
E{p^2}-xEp
&
yEp-Ep^2
\end{array}
\right],
\end{align*}
where we have defined $Ep = E[p | x \leq p \leq y]$. 

 It is easy to check that $Ep-x, y-Ep$ and $Ep^2 - xEp$ are positive since $x \leq p \leq y$. For the final term, observe that $yEp - p^2 = E(yp-p^2) = Ep(y-p) \geq 0$. We note that $Ep^2 = (Ep)^2 + \sigma^2$ where $\sigma^2$ is the variance of $p$ conditioned on $x \leq p \leq y$, and therefore can compute the determinant to be $-(y-x)\sigma^2$. 
 
Therefore, provided that $y>x$ and $F$ is non-degenerate, the matrix is invertible; which in turn implies that we can use Newton's method to find the solution. 
\section{Proof of Continuity -- $L_2$}
\label{sec:l2proof}

Given the (Lebesgue) integral in the objective, $a$ is not assumed continuous {\it a priori\/}.  We show that the optimal solution, however, is.  Let $t$ be the average target spend {\it per unit}.
\BEQ 
\begin{array}{cl} 
\inf_{a(\cdot)} & \frac{s}{2} \int (\frac{a(p)}{s} - \frac{d}{s})^2 ~ f(p)dp \\
\mbox{s.t.} & \int a(p)f(p)dp = d \\
& \int pa(p)f(p)dp \leq td \\
& 0 \leq \frac{a(p)}{s} \leq 1. 
\end{array} 
\EEQ
\noindent
We will ignore the nonnegativity constraint for simplicity. The Lagrangian is
\BEAS
L &=&\frac{1}{2s} \int \left(a(p) - d\right)^2 ~ f(p)dp + \int \mu(p)[a(p)-s]f(p) dp \\
&& ~~+~ z \left(\int a(p)f(p) dp - d\right)  \\
&& ~~+~ \lambda \left(\int pa(p)f(p)dp - td\right)
\EEAS
with $\mu(\cdot) \geq 0$ and $\lambda \geq 0$.  Note that $\mu(p) = 0$ if $a(p) < s$. By  Euler-Lagrange, 
$$ \frac{a}{s} - \frac{d}{s}+ \mu(p) + z + \lambda p = 0$$ Then,
$$\int a(p) f(p) dp = d ~ \Rightarrow ~ z =  \int (-\mu(p)-\lambda p) f(p) dp$$ and $$a(p) = d - sz -s \mu(p)- s\lambda p ~~~ \forall p$$

\medskip
\noindent
Now suppose there is a $p'$ such that $a(p') < s$. Then, $\mu(p') = 0$ and $$a(p') = d - sz - s\lambda p' < s$$ Then, for $p > p'$, we have
\BEAS
a(p) &=& d -sz -s\lambda p - s\mu(p) \\
~&\leq& d - sz -s \lambda p ~~ (\mbox{because} ~ \mu(p) \geq 0) \\
~ &<& d -sz -s \lambda p' = a(p')
\EEAS
Thus, $a(p)$ is monotone non-increasing. 

\medskip
\noindent
Let $$p_0 := \inf \{p \geq 0 ~:~ a(p) < s\}$$ Note that $p_0 \in [0, F^{-1}(d/s)]$. Then, 
$$
\begin{array}{llll}
a(p) & = & s & p < p_0 \\
& = & d -sz -s \lambda p &  0 \leq p_0 \leq p \leq p_m \\
& = & 0 & p_m \leq  p
\end{array}
$$
\noindent
Here, $p_m$ is such that $d - sz - s \lambda p_m = 0$. We now express $z$ in terms of $\lambda, p_0, p_m$ rather than in $\lambda, \mu(\cdot)$:
$$ z(\lambda, p_0, p_m) = \frac{sF(p_0) + d (F(p_m) - F(p_0) - 1) - s\int_{p_0}^{p_m}  \lambda p f(p) dp}{ s(F(p_m) -  F(p_0))}$$ Similarly for the Lagrangian:
\BEAS
L(\lambda, p_0) &=& \frac{(s-d)^2F(p_0)}{2s} + \int_0^{p_0} \lambda psf(p) dp \\
&& \frac{1}{2s} \int_{p_0}^{p_m} [s^2z^2 - s^2\lambda^2p^2+2s\lambda p d]f(p) dp 
\EEAS
It can be verified that 
$$\frac{\partial z}{\partial p_0} = \frac{f(p_0)(s - a(p_0))}{s(F(p_m)-F(p_0))}$$
$$\frac{\partial L}{\partial p_0} = \frac{f(p_0)}{2s}(s - a(p_0)^2 \geq 0$$
We see that either the optimum with respect to $p_0$ is achieved on the boundary ($p_0 = 0$ or $p_0 = F^{-1}(d/s)$) or that $a(p_0) = s$ at the optimum. We are not interested in the trivial case $p_0 = F^{-1}(d/s)$.  We thus have two cases:

\medskip
\noindent
{\bf Case 1.}  $$a(p) = d -sz -s\lambda p < s ~~~~ \forall p \geq 0$$

\noindent
{\bf Case 2.}
$$
\begin{array}{llll}
a(p) & = & s & p < p_0 \\
& = & s(1 + \lambda (p_0-p) & 0 \leq p_0 \leq p \leq p_m  \\
& = & 0 & p_m \leq p
\end{array}
$$We could combine the two cases by allowing $p_0$ to be negative.

\section{KL divergence}
\label{sec:klproof}
\label{sec:kl_appendix}

We want to minimize the KL divergence between $a(p)f(p)/d$ and $f(p)$:

$$\int \frac{a(p)f(p)}{d} \log \frac{a(p)f(p)/d}{f(p)} dp ~ = ~ \int f(p) \frac{a(p)}{d} \log \frac{a(p)}{d} dp$$ which is equivalent to minimizing 
$$\int f(p) a(p)\log a(p) dp$$ Given the (Lebesgue) integral in the objective, $a$ is not assumed continuous {\it a priori\/}.  We show that the optimal solution, however, is.  Let $t$ be the average target spend {\it per unit}.
Thus we have
\BEQ 
\begin{array}{cl} 
\inf_{a(\cdot)} & \int a(p)\log a(p) ~ f(p)dp \\
\mbox{s.t.} & \int a(p)f(p)dp = d \\
& \int pa(p)f(p)dp \leq td \\
& 0 \leq \frac{a(p)}{s} \leq 1. 
\end{array} 
\EEQ

Here, $t$ is the average target spend per unit. For feasibility, $t \geq \underbar{p} := \int_0^{F^{-1}(d/s)} psf(p) dp$. If $t \geq \bar{p} := \int_0^\infty p f(p) dp$, the optimal solution is $a(p) \equiv d$.  

\noindent
The Lagrangian is 
\BEAS
L &=&\int a(p) \log a(p) ~ f(p)dp + \int \mu(p)[a(p)-s]f(p) dp \\
&& ~~+~ \gamma\left(\int a(p)f(p) dp - d\right) + \lambda\left(\int pa(p)f(p)dp - td\right)
\EEAS
 with $\mu(\cdot) \geq 0$ and $\lambda \geq 0$.  Note that $\mu(p) = 0$ if $a(p) < s$. By  Euler-Lagrange, 
$$1 + \log a + \mu(p) + \gamma + \lambda p = 0$$ which gives
$$a = e^{-\gamma - 1} e^{-\mu(p)-\lambda p}$$ Then,
$$\int a(p) f(p) dp = d ~ \Rightarrow ~ d = e^{-\gamma - 1} \int e^{-\mu(p)-\lambda p} f(p) dp$$ which leads to $$a(p) = \frac{d}{Z(\lambda, \mu(\cdot))} e^{-\mu(p)-\lambda p} ~~~ \forall p$$

\medskip
\noindent
Now suppose there is a $p'$ such that $a(p') < s$. Then, $\mu(p') = 0$ and $$a(p') = \frac{d}{z}e^{-\lambda p'} < s$$ Then, for $p > p'$, we have
\BEAS
a(p) &=& \frac{d}{z} e^{-\lambda p - \mu(p)} \leq \frac{d}{z}e^{-\lambda p} ~~ (\mbox{because} ~ \mu(p) \geq 0) \\
~&<& \frac{d}{z} e^{-\lambda p'} = a(p')
\EEAS
 Thus, $a(p)$ is monotone non-increasing. 

\medskip
\noindent
Let $$p_0 := \inf \{p \geq 0 ~:~ a(p) < s\}$$ Note that $p_0 \in [0, F^{-1}(d/s)]$. Then, 
$$
\begin{array}{llll}
a(p) & = & s & p < p_0 \\
& = & \frac{d}{z}e^{-\lambda p} & p \geq p_0
\end{array}
$$
\noindent
We now express $z$ in terms of $\lambda, p_0$ rather than in $\lambda, \mu(\cdot)$:
$$ z(\lambda, p_0) = \frac{d\int_{p_0}^\infty e^{-\lambda p} f(p) dp}{d - s F(p_0)}$$ Similarly for the Lagrangian:
\BEAS
L(\lambda, p_0) &=& F(p_0) s \log s + \log \frac{d}{z(\lambda, p_0)} \int_{p_0}^\infty a(p) f(p) dp \\
&&~~+~ \lambda \int_0^{p_0} psf(p) dp - \lambda t d
\EEAS
from which follows
$$\frac{\partial L}{\partial p_0} = -s f(p_0)[\log \frac{a(p_0)}{s}  - \frac{a(p)}{s} + 1] \geq 0$$ We see that either the optimum with respect to $p_0$ is achieved on the boundary ($p_0 = 0$ or $p_0 = F^{-1}(d/s)$) or that $a(p_0) = s$ at the optimum. We are not interested in the trivial case $p_0 = F^{-1}(d/s)$. 

We have two cases for the form of the solution in the KL case.

\medskip
\noindent
{\bf Case 1.}  $$a(p) = \frac{d}{z(\lambda)} e^{-\lambda p} < s ~~~~ \forall p \geq 0$$

\noindent
{\bf Case 2.}
$$
\begin{array}{llll}
a(p) & = & s & p < p_0 \\
& = & se^{\lambda (p_0-p)} & p \geq p_0 \geq 0
\end{array}
$$

\begin{figure}[th]
\begin{center}
\includegraphics[bb=0 0 2.5in 2.5in]{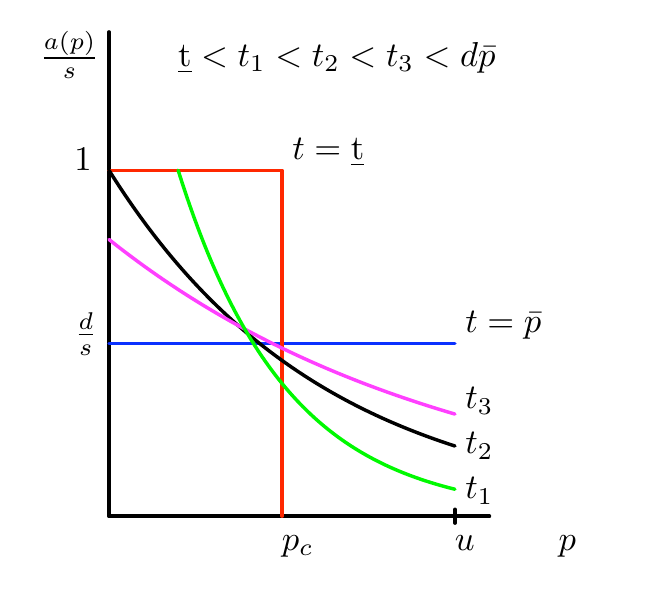}
\end{center}
\caption{Effect of target spend on KL-optimal allocation}
\label{fig:klalloc}
\end{figure}

\noindent
Figure \ref{fig:klalloc} shows the effect of varying target spend on the optimal allocation.

{\it Parametric supply distributions}
As an illustration, we consider the case when the supply distribution
$f$ is exponential: $f(p) = \gamma e^{-\gamma p}$.  Note that $\bar{p}
= \frac{1}{\gamma}$. As the budget decreases, a transition from Case 1
to Case 2 occurs at a certain budget. Until then, $a(p) = \frac{d}{z}
e^{-\lambda p}$ where $\frac{d}{z} \leq s,$ with equality at the
transitional budget. Demand constraint
$$\int \gamma \frac{d}{z} e^{-(\lambda + \gamma)p} dp = d$$ gives $$z
= \frac{\gamma}{\lambda + \gamma}$$ At the optimum, spend equals
budget: $$td = \int \gamma p
\frac{d(\lambda+\gamma)}{\gamma}e^{-(\lambda+\gamma)p}dp =
\frac{d}{\lambda + \gamma}$$ which leads to $\lambda_* = \frac{1}{t} -
\gamma$ and $z = \gamma t$.  Again, note that until the transition
happens, $\frac{d}{\gamma t} \leq s$, that is, $t \geq
\frac{d}{s\gamma} = \frac{d}{s}\bar{p}$. The optimal KL-divergence for
$\frac{d}{s}\bar{p} \leq t \leq \bar{p}$ is given by
$$\mbox{KL}_{\mbox{\sc opt}} = \gamma t - 1 - \log \gamma t$$ which is
0 when $t = \bar{p}$ and is $\frac{d}{s} - 1 -\log \frac{d}{s}$ at the
transitional budget.

\end{document}